\documentclass[journal,twoside, web]{ieeecolor}

\usepackage{etoolbox}
\makeatletter
\@ifundefined{color@begingroup}%
{\let\color@begingroup\relax
\let\color@endgroup\relax}{}%
\def\fix@ieeecolor@hbox#1{%
\hbox{\color@begingroup#1\color@endgroup}}
\patchcmd\@makecaption{\hbox}{\fix@ieeecolor@hbox}{}{\FAILED}
\patchcmd\@makecaption{\hbox}{\fix@ieeecolor@hbox}{}{\FAILED}

\usepackage{arxiv}
\usepackage{cite}
\usepackage{amsmath,amssymb,amsfonts}
\usepackage{algorithm}
\usepackage{algpseudocode}
\usepackage{graphicx}
\usepackage{textcomp}
\usepackage{booktabs}
\usepackage{multirow}
\usepackage{subfigure}
\usepackage{marvosym}
\usepackage{color}
\usepackage{stfloats}
\usepackage{xcolor}
\usepackage{ulem}
\usepackage{comment}
\usepackage{ulem}
\usepackage{stfloats}
\usepackage{threeparttable}

\usepackage[colorlinks,
            linkcolor=blue,
            anchorcolor=blue,
            urlcolor = blue,
            citecolor=blue
            ]{hyperref}
            
\algnewcommand{\LineComment}[1]{\State \(\triangleright\) #1}

\def\BibTeX{{\rm B\kern-.05em{\sc i\kern-.025em b}\kern-.08em
    T\kern-.1667em\lower.7ex\hbox{E}\kern-.125emX}}
\begin{document}
\title{ImplicitCell: Resolution Cell Modeling of \\Joint Implicit Volume Reconstruction and \\Pose Refinement in Freehand 3D Ultrasound} %
\author{Sheng Song, Yiting Chen, Duo Xu, Songhan Ge, Yunqian Huang, Junni Shi, Man Chen, Hongbo Chen, and Rui Zheng
\thanks{This work was supported by the Natural Science Foundation of
China (NSFC) under Grant 12074258. (Corresponding author: Hongbo Chen and Rui Zheng)}
\thanks{Sheng Song, Yiting Chen, Duo Xu, Songhan Ge, and Hongbo Chen are with School 
of Information Science and Technology, ShanghaiTech University, 
Shanghai, China. (e-mail: chenhb063@gmail.com)}
\thanks{Yunqian Huang, Junni Shi and Man Chen are with Tongren Hospital, Shanghai 
Jiao Tong University School of Medicine, Shanghai, China.}
\thanks{Dr. Rui Zheng is with School of Information Science and Technology, 
Shanghai Engineering Research Center of Energy Efficient and Custom 
AI IC, ShanghaiTech University, Shanghai, China (e-mail: zhengrui@shanghaitech.edu.cn)
}}

\maketitle

\begin{abstract}
Freehand 3D ultrasound enables volumetric imaging by tracking a conventional ultrasound probe during freehand scanning, offering enriched spatial information that improves clinical diagnosis. However, the quality of reconstructed volumes is often compromised by tracking system noise and irregular probe movements, leading to artifacts in the final reconstruction. To address these challenges, we propose ImplicitCell, a novel framework that integrates Implicit Neural Representation (INR) with an ultrasound resolution cell model for joint optimization of volume reconstruction and pose refinement. Three distinct datasets are used for comprehensive validation, including phantom, common carotid artery, and carotid atherosclerosis. Experimental results demonstrate that ImplicitCell significantly reduces reconstruction artifacts and improves volume quality compared to existing methods, particularly in challenging scenarios with noisy tracking data. These improvements enhance the clinical utility of freehand 3D ultrasound by providing more reliable and precise diagnostic information.

\end{abstract}


\section{Introduction}
\label{sec:introduction}

\subsection{Motivation}

\IEEEPARstart{U}{ltrasound} (US) imaging is a widely-used diagnostic tool in clinical practice due to its real-time capability and non-invasive nature. Compared to conventional 2DUS, 3DUS provides comprehensive volumetric visualization and enables quantitative analysis of anatomical structures, making it particularly valuable in clinic \cite{mohamed2019survey}.  
Freehand 3DUS, which combines a 1D US probe with a tracking device, offers a cost-effective and flexible solution for 3D volume acquisition \cite{fenster20233d}. In this solution, each 2D B-mode image is associated with its 6 Degree-of-Freedom (DoF) spatial pose, including position and orientation. The pose signal is 
typically collected by optical or electromagnetic (EM) tracking devices.
For point-of-care ultrasound (POCUS) applications - particularly in scenarios demanding portability and rapid deployment, EM tracking systems offer unique technical merits, facilitated by their non-line-of-sight operation and flexible sensor integration \cite{nguyen2024reliability, chen2020compact}. %
 
However, the spatial fidelity of freehand 3DUS reconstructions is often degraded by measurement noise inherent in tracking devices, particularly EM systems \cite{esposito2019total}, and further compounded by involuntary hand movements (e.g., jitter) during scanning \cite{9869860}. These noise sources introduce substantial inaccuracies in pose estimation, which can significantly compromise reconstruction quality and reduce diagnostic precision, especially in high-resolution applications like Carotid Atherosclerosis (CA) diagnosis \cite{johri2020recommendations}.
Although existing methods attempt to mitigate these artifacts through explicit pose refinement with priors (e.g., smoothness), their effectiveness is limited by the complex and unpredictable nature of freehand probe movements\cite{mozaffari2017freehand, liang2022spatiotemporal}.

Implicit neural representation (INR), which utilizes multilayer perceptrons (MLP) with position encoding (PE), has emerged as a powerful technique for representing complex 3D structures across medical imaging tasks \cite{xu2023nesvor,wu2023joint, 10793769}. For US imaging, INR has demonstrated success over traditional discrete methods in both 2DUS synthesis and 3DUS reconstruction, with its continuous representation being advantageous for freehand 3DUS where data is acquired as irregularly sampled slices in 3D space \cite{wysocki2023ultra,song2022development,yeung2021implicitvol}.
While INR has shown promise for US representation, existing methods often overlook the fundamental physics of ultrasound scatterer interaction within tissues, which is essential for addressing pose uncertainty and reconstruction artifacts in freehand 3DUS. Specifically, the formation of pixel intensities through coherent summation of backscattered echoes within resolution cells provides inherent spatial correlations, yet this physics-based insight remains unexploited in current approaches.

In this study, we propose ImplicitCell, a physics-aware INR model that incorporates US backscattering through an explicit resolution cell model to simultaneously optimize volume reconstruction and pose refinement. By capturing the spatial correlations of backscattered echoes in overlapping resolution cells, ImplicitCell enables high-quality 3D reconstruction from POCUS acquisitions, demonstrating robust performance even with noisy EM tracking data and complex motion patterns.

\subsection{Related Works}

\subsubsection{Freehand 3DUS Volume Reconstruction}

Freehand 3DUS reconstruction aims to derive a 3D volume from a sequence of 2DUS images with corresponding pose signals. Current methods, primarily based on explicit discrete models, focus on two main objectives: computational efficiency and reconstruction quality. For efficiency improvement, e.g., Victoria \textit{et al.} \cite{victoria2023real} utilized octrees for real-time voxel querying, and Chen \textit{et al.} \cite{chen2021uffc} accelerated traditional algorithms such as voxel nearest neighbor (VNN) and pixel nearest neighbor (PNN) through optimized numerical operations. For quality enhancement, e.g., Moon \textit{et al.} \cite{moon20163d} employed a piecewise smooth Markov random field to preserve sharp edges while reducing noise, and Wang \textit{et al.} \cite{wang2023adaptive} achieved high-quality reconstruction through adaptive tetrahedral interpolation of point clouds. However, these methods heavily rely on either high-precision tracking systems or sophisticated pose refinement algorithms to ensure reconstruction precision.

\subsubsection{Pose Signal Processing}
Pose signal processing is vital for improving tracking accuracy in medical imaging, especially for noise-susceptible EM systems. Filter-based sensor fusion techniques have demonstrated significant effectiveness for such task. Lang \textit{et al.} \cite{lang2009fusion} combined ultrasound speckle correlation with unscented Kalman filtering (UKF) to reduce metallic distortion artifacts. Zhang \textit{et al.} \cite{zhang2021research} integrated inertial measurement units (IMU) with extended Kalman filter (EKF) for accuracy high-frequency motion capture. Besides Kalman filters, Luo \textit{et al.} \cite{luo2023constrained} developed a Constrained Evolutionary Diffusion Filter (CEDF) that fuses EM tracking with computed tomography (CT) and endoscopic imaging to enhance robustness in complex scenarios.
Beyond filters, Esposito \textit{et al.} \cite{esposito2019total} first explored convex optimization-based pose denoising for high-quality 3DUS volume reconstruction. The pose signals obtained from tracking devices are regarded as \textit{manifold-valued signals} and optimized by applying total variance regularization (TVR). The proposed method can effectively reduce reconstructed artifacts while preserving trajectory fidelity, particularly beneficial for EM tracking systems.

Additionally, \textit{sensorless} approaches eliminate the need for external tracking devices by estimating poses directly from ultrasound image features. These methods pioneered the use of resolution cell modeling to analyze speckle patterns for pose estimation \cite{gee2006sensorless, afsham2015nonlocal}, resulting in various speckle decorrelation-based algorithms \cite{app14177991}. Recent deep learning approaches have shown significant progress by training neural networks on large-scale data \cite{prevost20183d,luo2023recon}. However, these methods still face challenges with accuracy and spatial drift compared to tracked methods \cite{fenster20233d}.

\subsubsection{Implicit Neural Representation}

INR models have demonstrated significant success in various 3D vision tasks \cite{mildenhall2021nerf,lin2021barf} and emerged as powerful tools for medical image reconstruction across modalities, effectively addressing challenges such as motion artifacts \cite{xu2023nesvor,wu2023joint,dagli2024nerf}. In the context of freehand 3DUS reconstruction, these approaches have shown particular promise in improving volume quality. 
A common paradigm of INR-based freehand 3DUS reconstruction methods directly maps 3D coordinates to intensity values by training networks with 3D pixel positions (derived from 2D US images and pose data) and their corresponding intensities \cite{song2022development, velikova2024implicit,wysocki2023ultra, yeung2021implicitvol}.
Upon this foundation, several researches explored the integration of semantic information with INR models to enhance the delineation of arteries \cite{song2022development, velikova2024implicit}. Ultra-NeRF pioneered the integration of US wave propagation physics into INR models, to achieve view-dependent and physically plausible ultrasound image synthesis \cite{wysocki2023ultra}, and Nerf-US further demonstrated effective motion artifact reduction in clinical US image synthesis \cite{dagli2024nerf}. ImplicitVol \cite{yeung2021implicitvol}, which optimizes INR model and slice transformations simultaneously, effectively compensates for fetal motion between US slices. 
Recently, some studies have been using INR for surface and shape reconstruction. However, these types of algorithms require segmented masks, which are not considered in this paper \cite{velikova2024implicit, chen2024neural}.

\subsection{Contribution}%
This paper introduces \textbf{ImplicitCell}, a robust framework for reconstruction artifact reduction in freehand 3DUS that integrates resolution cell modeling with INR, optimized specifically for POCUS systems with noisy EM tracking. This paper features the following contributions: 1) We propose a novel resolution cell model that divides each pixel's ultrasound resolution cell into subcells, enabling accurate modeling of spatial overlap between adjacent US pixels. 2) We develop a unified framework that leverages such physics-based modeling with INR for joint optimization of volume reconstruction and pose refinement, achieving robust results under noisy pose signals from EM tracking systems. 3) We design several regularization techniques to counteract drifting and noise specific to freehand 3DUS, further improving volume quality and pose accuracy. 4) We perform various experiments with an EM-tracked POCUS system using phantom, volunteer and clinical datasets. The results demonstrate the effectiveness of ImplicitCell in reducing reconstruction artifacts and improving volume quality compared to existing methods. The source code and phantom dataset %
will be released at \url{https://github.com/Vilour/ImplicitCell}.

\section{Methodology}
\begin{figure*}[h]
    \centerline{\includegraphics[width=1.9\columnwidth]{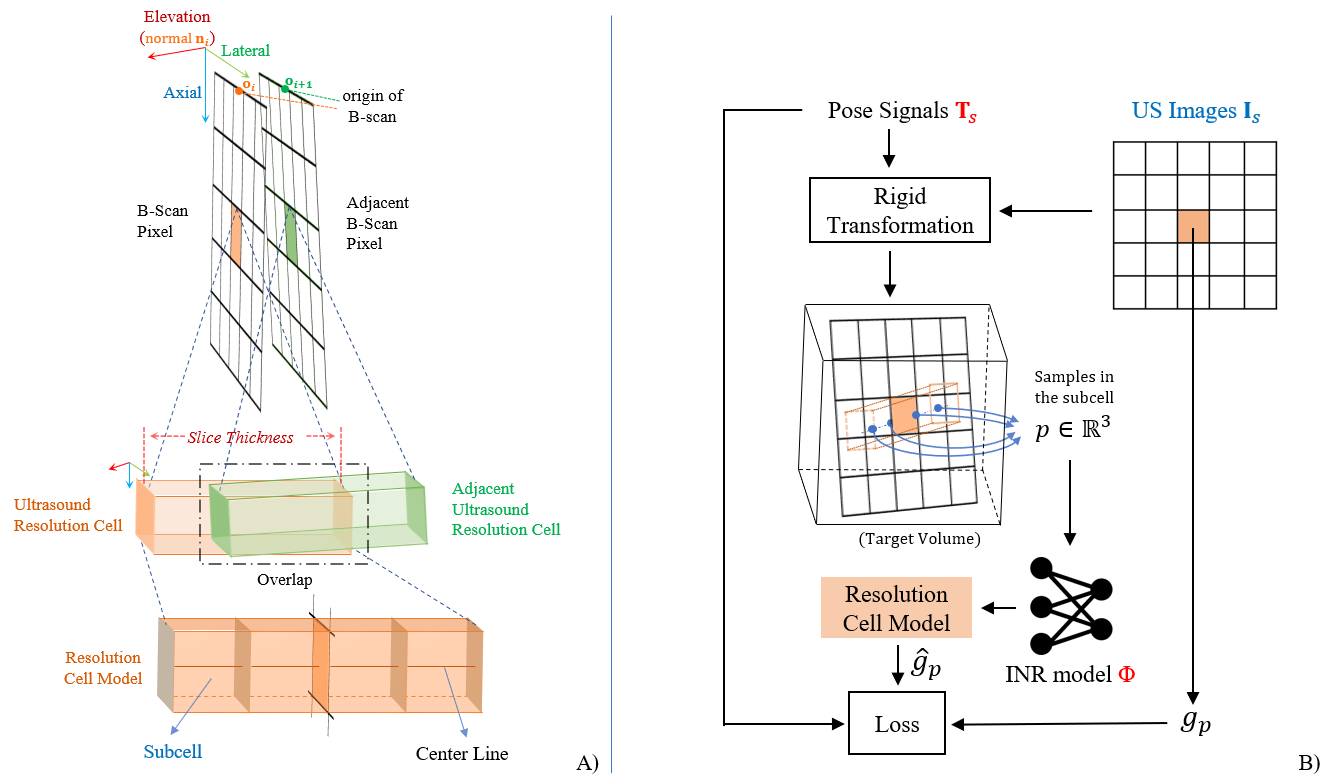}}
    \caption{A visual overview of the ImplicitCell framework and its key components. A). Two pixels on adjacent frames and their corresponding US resolution cells. Each resolution cell is devided into subcells in our resolution cell model. B). The overview of our proposed method. The trainable parameters are highlighted in \textcolor{red}{red}. }
    \label{fig::pipeline}
\end{figure*}
\label{sec:methodology}

In this section, we first introduce the fundamentals of freehand 3DUS imaging and the challenges introduced by pose signal noise. We then formulate the ultrasound resolution cell model and present the ImplicitCell framework, which integrates this model with INR for joint optimization of volume reconstruction and pose refinement (Fig. \ref{fig::pipeline}B). Finally, we detail the optimization process and training strategy for ImplicitCell.
\subsection{Preliminaries}
\subsubsection{Freehand 3DUS Imaging}
Freehand 3DUS imaging involves acquiring a series of B-mode US images $\mathbf{I}_s = \{ I_i : i = 1 \dots N, I_i \in \mathbb{R}^{W \times H} \}$ with corresponding pose signals $\mathbf{T}_s = \{ T_i : i = 1 \dots N, T_i \in \text{SE}(3) \}$, where $W$ and $H$ represent the width and height of the images, respectively, $N$ is the total number of images acquired, $\text{SE}(3)$ represents the special Euclidean group that includes rotations ($R \in \text{SO}(3)$) and translations ($t \in \mathbb{R}^3$) \cite{esposito2019total}. The 3D location of US pixel $p_{i,w,h}\in \mathbb{R}^3$ is calculated using a rigid transformation function $G\left(w,h,T_i, d_{\text{pixel}} \right)$, where $w$ and $h$ are the pixel coordinates, $T_i$ is the pose signal, and $d_{\text{pixel}}$ is the pixel size of US images. Denote the grey scale intensity of pixel as $g_{p}$, and the 3DUS volume $V\in \mathbb{R}^{V_l \times V_w \times V_h} $ is reconstructed from $\left\{ \left( p_m, g_{p_m} \right) \right\}^M_{m=1}$, where $M= I \times W \times H $. $V_l$, $V_w$, and $V_h$ represent the length, width, and height of the volume, respectively. %

\subsubsection{Pose Signal Noise}
In freehand 3DUS reconstruction, each voxel value is determined by the contributions from nearby ultrasound pixels $P_v \subseteq  \left\{ \left( p_m, g_{p_m} \right) \right\}^M_{m=1}$. When pose measurements contain errors, pixels may be incorrectly mapped to $P_v$, leading to inconsistent intensity values within local neighborhoods of the voxel. These inconsistencies manifest as spatial artifacts in the reconstructed volume. To reduce such artifacts, we need to minimize these local intensity inconsistencies through pose refinement, which requires understanding the spatial relationship between pixels in 3D space.

\subsection{Ultrasound Resolution Cell}
\label{sec::cell_model}
To model the spatial correlation of pixels, we leverage the fact that the beam from US probe has the physical thickness (usually several millimeters wide), resulting in a resolution region rather than a single pixel \cite{scholten2023differences}. The width of the beam is known as \textit{slice thickness}, and the resolution region is known as \textit{resolution cell}. 
The intensity of US images $g_p$ is proportional to the total scatters from \textit{resolution cell} \cite{prager2003sensorless}. Denote the echo intensity of scatters at $p\in \mathbb{R}^3$ as $g_{\text{scatter}}\left( p \right)$, and the intensity of pixel $g_p$ can be expressed as the integral of all scatters over the \textit{resolution cell}:
\begin{equation}
    g_p = \int _{\text{Cell}} g_{\text{scatter}}\left( p \right) dp
    \label{eq::physical_}
\end{equation}
The size of the \textit{resolution cell} is determined by pixel size $d_{\text{pixel}}$ and \textit{slice thickness}. \textit{Slice thickness} is usually in the millimeter range \cite{scholten2023differences}, resulting in a large \textit{resolution cell} that overlaps with neighboring cells (see Fig. \ref{fig::pipeline}A). The shared scatters in overlapping area lead to similarity between pixels and their neighbors, which can be expolited for spatial correlation modeling. Furthermore, we devide the \textit{resolution cell} into $K$ \textit{subcells} along the center line, and the intensity of pixel $g_p$ can be modeled as the summation of all \textit{subcell} intensities:
\begin{equation}
    g_p =  \sum_{k=1}^{K} \int _{\text{Subcell}_k} g_{\text{scatter}}\left( p \right) dp
    \label{eq::subcell}
\end{equation}
Thereby, the shared scatters in overlapping area can be modeled as the shared scatters in \textit{subcells}, which can be used to characterize the spatial correlation among pixels. Moreover, scatters closer to image plane contribute more to the pixel intensity, making the Gaussian distribution a popular choice for modeling scatters in \textit{resolution cell} \cite{prager2003sensorless, gee2006sensorless}. Similarly, we apply a discrete Gaussian function $\eta$ to perform a weighted summation of \textit{subcell} intensities in Eq. \ref{eq::subcell}:
\begin{equation}
    g_p = \sum_{k=1}^{K} \eta\left( d_k \right) \int _{\text{Subcell}_k} g_{\text{scatter}}\left( p \right) dp
    \label{eq::gaussian_sc}
\end{equation}
where $d_k$ is the distance of \textit{subcell} to the image plane. The Gaussian function $\eta$ is defined as:
\begin{equation}
    \begin{split}
        & \eta\left( d_k \right) = \frac{1}{Z} \exp \left(-\frac{(d_k)^2}{2 \sigma^2}\right) \\
        & Z=\sum\limits_{k=1}^K \exp \left(-\frac{(d_k)^2}{2 \sigma^2}\right)
    \end{split}
\end{equation}
where $\sigma$ regulates the declining influence of scatters as distance increases. 

\subsection{Joint INR Modeling and Pose Refinement}
\subsubsection{INR with US Resolution Cell}
INR models represent 3D scenes through neural networks by predicting values at arbitrary spatial coordinates. This implicit representation naturally aligns with modeling scatter echo intensities in Eq. \ref{eq::physical_}. Specifically, we denote our INR model as \( \Phi: \mathbb{R}^{3} \rightarrow \mathbb{R} \), where the output of $\phi$ represents the echo intensity of scatters, approximating $g_{\text{scatter}}$. Moreover, instead of direct computation of the integral on \textit{subcells} in Eq. \ref{eq::subcell}, we approximate it with Monte Carlo sampling. We model the spatial domain of \textit{subcells} as a random variable $X_{\text{sc}}$, where $X_{\text{sc}}$ takes values in the \textit{subcell} region. 
Note that the pixel size $d_{\text{pixel}}$ of US image is always several times smaller than \textit{slice thickness} \cite{scholten2023differences}, and therefore we can approximate the region of \textit{subcell} as a segment of center line. Denote the \textit{slice thickness} as $S_{t}$, the normal vector of image plane as $\mathbf{n}$, and the center of \textit{subcell} as $c_{\text{sc}}\in \mathbb{R}^3$, and we define the random variable $X_{\text{sc}}$ as:
\begin{equation}
    X_{\text{sc}} \thicksim c_{\text{sc}} + \mathbf{n} \cdot U \left( -\frac{S_t}{2K}, \frac{S_t}{2K} \right)
\end{equation}
Therefore, we draw $N_s$ samples from $X_{\text{sc}}$, and the intensity of pixel $g_p$ can be approximated as:
\begin{equation}
    \hat{g}_p = \sum_{k=1}^{K} \eta\left( d_k \right)  \left( \frac{1}{N_s} \sum_{n=1}^{N_s} \Phi\left( p_{\text{sc},k}^{(n)}\right) \right)
    \label{eq::prediction}
\end{equation}

\subsubsection{Pose Signal Refinement}
\label{sec::pose_refinement}
The resolution cell and INR models work together to capture spatial correlations between pixels, enabling pose refinement by minimizing intensity inconsistencies. For efficient pose refinement, we decouple pose signals $T_i$ into rotation $SO(3)$ and translation $T(3)$ components following \cite{lin2023parallel}. This allows us to parameterize each pose signal $T_i$ with a vector $v_T^i \in \mathfrak{so}(3) \times \mathfrak{t}(3) \cong \mathbb{R}^6$. Our parameterization ensures pose updates follow straight lines in translation space and geodesic paths on the rotational manifold, leading to stable and efficient optimization.
\ 
\newline
\indent Refining pose signals without constraints may lead to suboptimal solutions, i.e., similar images tend to cluster together, while adjacent images with noisy pose may drift apart, resulting in significant artifacts in the reconstructed volume.
To mitigate this issue, we constrain pose optimization within physically plausible boundaries. Specifically, during freehand 3DUS acquisition, the high frame rate (20-60 fps) relative to slow probe movement ensures small and consistent pose differences between consecutive frames. Therefore, we implement a sliding window approach to constrain pose optimization within reasonable ranges of neighboring frames. Since optimization uncertainty is highest along the elevation axis of images, we primarily constrain pose refinement in this direction. To do this, we define a metric $\mathfrak{D}$ that calculates the perpendicular (normal) distance between adjacent US image planes. Specifically, we project the vector connecting the origins of two consecutive frames onto the normal direction of the first frame, as illustrated in Fig. \ref{fig::pipeline}A:
\begin{equation}
\label{eq::g}
\mathfrak{D}_{i} := \vert \left( \mathbf{o}_{i+1} - \mathbf{o}_{i} \right) \cdot \mathbf{n}_{i} \vert
\end{equation}
Here, $\mathbf{o}$ and $\mathbf{n}$, represent the origin and the normal of the US image plane respectively. 
Given the pose signals $\mathbf{T}_s$, we can calculate all the adjacent projection as $\mathcal{D}_{\left[ 1,N \right]}^s :=\left\{ \mathfrak{D}_{i}\right\}_{i=1\ldots N-1}$. Denote the length of sliding window as $W_s$, we have the optimization interval of $\mathfrak{D}_{i}$ as:
\begin{equation}
    \mathfrak{D}_{i} \in \left[\min \mathcal{D}_{\left[ i-W_s,i+W_s \right]}^s, \max \mathcal{D}_{\left[ i-W_s,i+W_s \right]}^s \right]
    \label{eq::t_constraint}
\end{equation}
The bounds placed on $\mathfrak{D}_{i}$ restrict the allowable translational displacement between adjacent pose signals in 3D space. %
Similarly, to constrain the orientation components, we denote the angle between adjacent poses as $\theta$, and 
all the adjacent angles as $\varTheta_{\left[ 1,N \right]}^s :=\left\{ \theta_{i}\right\}_{i=1\ldots N-1}$.
The optimization interval of $\theta_{i}$ is denoted as:
\begin{equation}
    \theta_{i} \in \left[\min \varTheta _{\left[ i-W_s,i+W_s \right]}^s, \max \varTheta_{\left[ i-W_s,i+W_s \right]}^s \right]
    \label{eq::r_constraint}
\end{equation}
Based on the constrains, we construct the regularization term to penalize the deviation of pose signals from their optimization intervals (Sec. \ref{sec::pose_reg}).

\subsection{Model Optimization}
The joint optimization problem of 3DUS volume reconstruction and pose signal refinement can be formulated as:
\begin{equation}
    \mathbf{T}_s^*, \Phi^* = \underset{\mathbf{T}_s, \Phi}{\operatorname{argmin}} \chi \left( \mathbf{T}_s, \mathbf{I}_s, \Phi\right)
    \label{eq::optimization_function_}
\end{equation}
where $\chi$ is the objective function, which consists of intensity loss and regularization terms. 

\subsubsection{Intensity Loss}
During training, we optimize both the INR model and pose parameters by minimizing the difference between predicted intensities (Eq. \ref{eq::prediction}) and observed intensities from the original ultrasound images. We compute this intensity loss using mean squared error (MSE):
\begin{equation}
    l_{i} = \frac{1}{B} \left\lVert  g_p - \hat{g}_p \right\rVert^2_2  
\end{equation}
where $B$ is the number of sampled pixels, $g_p$ represents the observed pixel intensity from US images, and $\hat{g}_p$ denotes the predicted intensity from our model (Eq. \ref{eq::prediction}).

\subsubsection{Pose Regularization}
\label{sec::pose_reg}
As described in Sec. \ref{sec::pose_refinement}, we penalize the deviation of pose signals from their optimization intervals using function $\varphi$:
\begin{equation}
    \varphi\left( x, a, b \right) := \left[ \max\left(a - x, x - b, 0 \right) \right]^2
\end{equation}
where $a$ and $b$ are the lower and upper bounds of the optimization interval. For each adjacent distance $\mathfrak{D}_{i}^*$ and angle $\theta_{i}^*$ in optimization, we can calculate the deviation as:
\begin{equation}
    \begin{split}
        & \varphi^i_{\mathcal{D}} = \varphi\left( \mathfrak{D}_{i}^*, \min \mathcal{D}_{\left[ i-W_s,i+W_s \right]}^s, \max \mathcal{D}_{\left[ i-W_s,i+W_s \right]}^s \right) \\
        & \varphi^i_{\theta} = \varphi\left( \theta_{i}^*, \min \varTheta_{\left[ i-W_s,i+W_s \right]}^s, \max \varTheta_{\left[ i-W_s,i+W_s \right]}^s\right) \\
    \end{split}
\end{equation}
Finally, the regularization term is defined as:
\begin{equation}
    \begin{split}
        & l_{\mathfrak{D}} := \max \left\{ \varphi^i_{\mathcal{D}} \right\}_{i=1\ldots N-1} \\
        & l_\theta := \max \left\{ \varphi^i_{\theta} \right\}_{i=1\ldots N-1} \\
    \end{split}
\end{equation}

\subsubsection{Volume Regularization}
\label{sec::volume_reg_}

Due to the ill-posed problem of solving Eq. \ref{eq::gaussian_sc}, we adopt first-order regularization method to improve the reconstruction quality. The total variance regularization is one of the most common regularizers, and represented as:
\begin{equation}
    R_V = \int|\nabla V(v)| dv
\end{equation}
Although the gradient of the volume $\nabla V(v_i)$ can be obtained through automatic differentiation, the computational cost will increase significantly. Therefore, we leverage existing variables in training stage to approximate $|\nabla V(v)|$ through the weighted variance of subcells. From Eq. \ref{eq::gaussian_sc}, we define the volume regularizer at sample $p$ as:
\begin{equation}
    R_V(p) = \sum\limits_{i=1}^K\left( \eta\left( d_{k}^i \right)\cdot \left( \frac{1}{N_s} \sum_{n=1}^{N_s} \Phi\left( p_{\text{sc},k}^{(n)} \right) - \hat{g}_p \right)^2 \right)    
\end{equation}

Combining these elements, the overall loss function for our optimization problem Eq. \ref{eq::optimization_function_} is defined as:
\begin{equation}
    \chi = l_{i} + \beta_{\mathfrak{D}} l_{\mathfrak{D}} + \beta_\theta l_\theta + \beta_R R_V
    \label{eq::loss_}
\end{equation}
Here, $\beta_{\mathfrak{D}}, \beta_\theta, \beta_R $ are hyperparameters that balance the contributions of each regularization component to the overall loss, ensuring optimal reconstruction performance.

Once trained, the INR model $\Phi^*$ implicitly defines the target volume $V$. We query $V$ by evaluating $\Phi^*$ at the voxel locations $\mathbf{v} = \left\{ v_1, v_2, \ldots, v_{V_l\times V_w \times V_h}, v\in \mathbb{R}^3 \right\}$:
\begin{equation}
    V = \left\{ \Phi^*\left( v_i \right) \right\}_{i=1\ldots V_l \times V_w \times V_h}
\end{equation}

\section{Experiments}
\subsection{Datasets}
We construct three datasets to evaluate our method: phantom dataset, common carotid artery (CCA) dataset, and CA dataset. We utilize a portable freehand 3DUS imaging system to acquire the US data. The system primarily consists of a 2D linear probe (Clarius, L738-K, Canada) for image acquisition and an EM tracking system (Polhemus, G4 unit, U.S.A) for pose signal acquisition. More details about the system can be found in \cite{li2023automatic}. 
\subsubsection{Phantom Dataset}
\begin{figure}[h]
    \centering
    \centerline{\includegraphics[width=0.9\columnwidth]{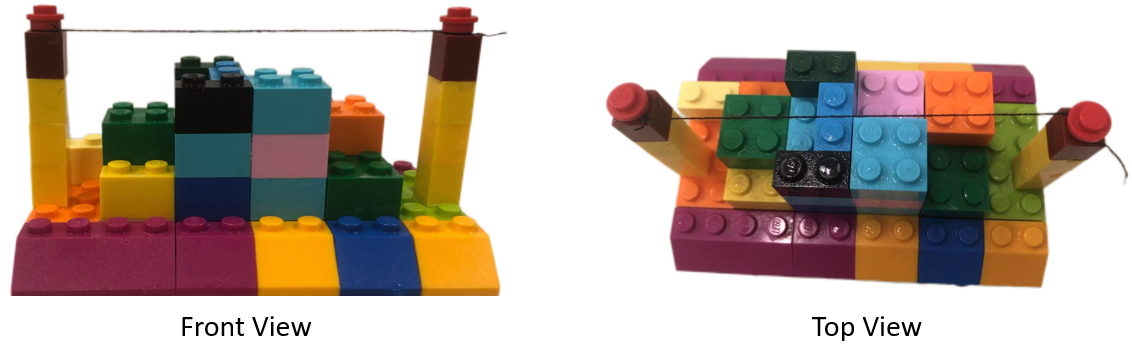}}
    \caption{Front view and top view of Lego phantom.}
    \label{fig::phantom}
\end{figure}

Our phantom dataset consists of a Lego brick construction with an embedded straight wire, as shown in Fig. \ref{fig::phantom}. To thoroughly validate reconstruction performance under realistic conditions, we acquired 29 scans of this phantom in a water tank environment. Based on the noise level of pose signals, these scans were categorized into two groups: heavy noise and light noise conditions.

\subsubsection{CCA Dataset}
This dataset contains 36 scans of the CCA from volunteers, with both the left and right CCA of each volunteer being scanned. Additionally, paired $T_2^*$-weighted magnetic resonance image (MRI) sequences for each volunteer were conducted for reference. Each reconstructed volume and MRI reference were manually segmented to extract the CCA structure \cite{liu2023exploring}.

\subsubsection{CA Dataset}
This is a clinic dataset collected from the local hospital, which contains 16 scans of CA. Each subject's scan includes an atherosclerotic plaque as a pathological feature. Additionally, corresponding longitudinal US images of CA are obtained for reference at the hospital using a clinical US diagnosis system (Siemens Healthineers, ACUSON Sequoia, Germany) equipped with a 10L4 transducer. Ethics approval was granted by the local health ethics board.

\subsection{Evaluation Metrics}
According to the feature of datasets, we employ various dataset-specific evaluation metrics to comprehensively assess the performance of methods.

\subsubsection{Phantom Dataset}
To evaluate phantom dataset reconstruction accuracy, we use the Line Fitting Error (\textbf{LFE}). LFE quantifies the deviation of a reconstructed straight wire from its ideal linear form. Inspired by line-based evaluations in volumetric ultrasound imaging \cite{lavenir2023tbme}, LFE is calculated by: 1) extracting the wire sub-volume, 2) computing barycenters of axial slices, 3) fitting a least-squares line to these points, and 4) calculating the mean distance from the barycenters to the fitted line.

\subsubsection{CCA Dataset}
We evaluate reconstruction quality using: 1) Centerline Distance (\textbf{CLD}) between reconstructed and MRI reference volumes for reconstruction accuracy \cite{acosta2017multi}, and 2) absolute Gaussian Curvature (\textbf{GC}) for surface smoothness assessment \cite{cohen2003restricted}.

\subsubsection{CA Dataset}
We assess reconstruction quality through: 1) image similarity metrics (Structural Similarity, \textbf{SSIM} \cite{wang2004image}, and Learned Perceptual Image Patch Similarity, \textbf{LPIPS} \cite{zhang2018unreasonable}) between aligned slice from reconstructed volume and longitudinal scans, and 2) Surface Irregularity Index (SII, rad/mm) \cite{kanber2013quantitative} and its difference (\textbf{$\Delta$SII}) between aligned slices and reference scans, serving as clinically established metrics for plaque morphology.

\subsubsection{Ablation Study}
For ablation studies on phantom dataset, we use \textbf{LFE} for geometric accuracy and Peak Signal-to-Noise Ratio (\textbf{PSNR}) for INR model performance assessment.

\subsection{Baseline}
We adopt both two traditional pipelines and two INR-based methods as baselines. The traditional pipeline includes volume reconstruction and pose refinement. For volume reconstruction, we employed Fast-dot-projection (FDP) with VNN \cite{chen2021uffc}, which directly projects pixels onto voxels. For pose refinement, we used TVR \cite{esposito2019total}, a state-of-the-art method for noise reduction in EM tracking systems, and we denote reconstructing from refined pose signals as FDP-TVR. For the INR-based method, we adopted INR model from Song \textit{et al}\cite{song2022development} and ImplicitVol \cite{yeung2021implicitvol}. The INR model contains a PE function and an 8-layer MLP, and the loss for semantic mask loss was disabled in our experiment\cite{song2022development}. For ImplicitVol, we replaced the initial poses from PlaneInVol \cite{yeung2021learning} predictions with pose signals from our tracking system.
\begin{figure*}[t]
    \centerline{\includegraphics[width=1.9\columnwidth]{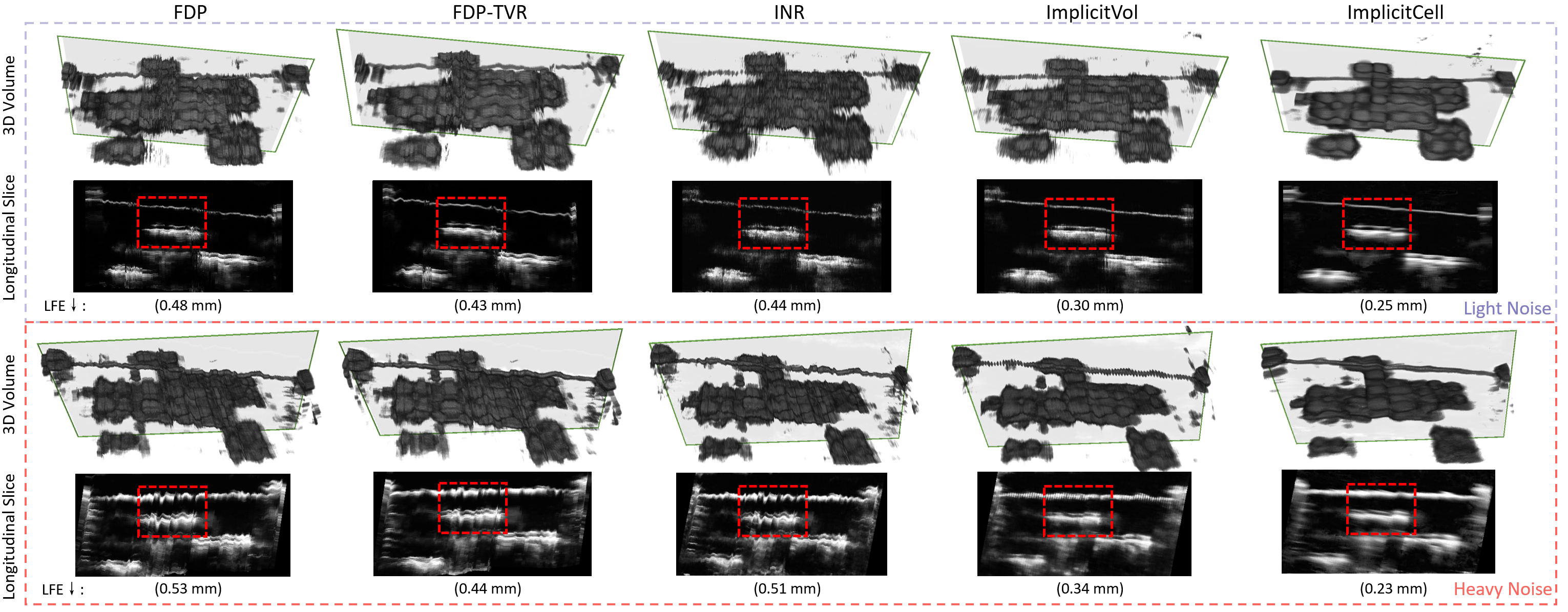}}
    \caption{Visual comparison of reconstruction quality across different methods under ``Light Noise" (top row) and ``Heavy Noise" (bottom row) conditions on phantom dataset. For each method, the top image shows the 3D rendering of the reconstructed volume, and the bottom image displays a longitudinal slice whose position is illustrated in the 3D view (green wireframe indicating the longitudinal slice position). The red dashed box highlights the differences. Numerical values below the slices indicate the LFE metric (mm, lower is better).}
    \label{fig::qualitative_phantom} %
\end{figure*}
\subsection{Implementation Details}
Our INR model $\Phi$ is defined as $\Phi = \gamma_{\text{hash}} \circ \phi$, where $\gamma_{\text{hash}}$ is the multiresolution \textit{hash encoding} function \cite{muller2022instant} and $\phi$ is the MLP. We employed a course-to-fine strategy for the resolution cell during training, starting from an initial number of \textit{subcells} $K_{\text{init}}=3$ and gradually increasing to the final setting $K=8$. Similarly, the scale level $L$ of the encoding function $\gamma_{\text{hash}}$ increases after each epoch, reaching the final setting at the last epoch.
The MLP $\phi$ uses SIREN \cite{sitzmann2020implicit} as the activation function, with 4 hidden layers of dimension 128. The slice thickness $S_t$ is set to 3.0 mm for the phantom dataset and 1.5 mm for the carotid dataset, according to the depth of US images. The weights in the loss function $\beta_{\mathfrak{D}}, \beta_{\theta}$, and $\beta_{R}$ are set to 0.03, 30, and 1, respectively. The voxel size of all reconstructed volumes is set to 0.2*0.2*0.2 mm.
Other implementation details can be found in our released code.
Our method was tested on an NVIDIA RTX 3090 GPU with 24 GB RAM. The model was implemented with PyTorch and Tiny CUDA NN.
\begin{table*}[tp]
    \centering
    \caption{Performance Evaluation of Different Methods across Three Datasets}
    \label{tab:performance_results}
    \begin{tabular}{l c c c c c c c c c c c c}
    \toprule
     & \multicolumn{2}{c}{Phantom Dataset} & \multicolumn{4}{c}{CCA Dataset} & \multicolumn{6}{c}{CA Dataset} \\
    \cmidrule(r){2-3} \cmidrule(r){4-7} \cmidrule(l){8-13}
           & \multicolumn{2}{c}{$\downarrow$ LFE (\textit{mm})} & \multicolumn{2}{c}{$\downarrow$ CLD (\textit{mm})} & \multicolumn{2}{c}{$\downarrow$ GC} & \multicolumn{2}{c}{$\uparrow$ SSIM} & \multicolumn{2}{c}{$\downarrow$ LPIPS} & \multicolumn{2}{c}{$\downarrow$ $\Delta$SII (\textit{rad/mm})} \\
    \cmidrule(r){2-3} \cmidrule(l){4-5} \cmidrule(l){6-7} \cmidrule(l){8-9} \cmidrule(l){10-11} \cmidrule(l){12-13}
    Method      & Mean & SD & Mean & SD & Mean & SD & Mean & SD & Mean & SD & Mean & SD \\
    \midrule
    FDP & \multicolumn{1}{|c}{0.56} & \multicolumn{1}{c}{0.23} & \multicolumn{1}{|c}{0.51} & \multicolumn{1}{c}{0.14} & \multicolumn{1}{|c}{0.32} & \multicolumn{1}{c}{0.03} & \multicolumn{1}{|c}{0.41} & \multicolumn{1}{c}{\textbf{0.10}} & \multicolumn{1}{|c}{0.30} & \multicolumn{1}{c}{0.08} & \multicolumn{1}{|c}{0.98} & \multicolumn{1}{c}{0.78} \\
    
    FDP-TVR & \multicolumn{1}{|c}{0.59} & \multicolumn{1}{c}{0.19} & \multicolumn{1}{|c}{0.51} & \multicolumn{1}{c}{0.14} & \multicolumn{1}{|c}{0.26} & \multicolumn{1}{c}{\textbf{0.02}} & \multicolumn{1}{|c}{0.42} & \multicolumn{1}{c}{0.11} & \multicolumn{1}{|c}{0.26} & \multicolumn{1}{c}{\textbf{0.07}} & \multicolumn{1}{|c}{0.45} & \multicolumn{1}{c}{0.31} \\
    
    INR & \multicolumn{1}{|c}{0.55} & \multicolumn{1}{c}{0.22} & \multicolumn{1}{|c}{0.50} & \multicolumn{1}{c}{0.12} & \multicolumn{1}{|c}{0.29} & \multicolumn{1}{c}{0.04} & \multicolumn{1}{|c}{0.43} & \multicolumn{1}{c}{\textbf{0.10}} & \multicolumn{1}{|c}{0.30} & \multicolumn{1}{c}{0.09} & \multicolumn{1}{|c}{0.87} & \multicolumn{1}{c}{0.72} \\
    
    ImplicitVol* & \multicolumn{1}{|c}{0.33} & \multicolumn{1}{c}{0.12} & \multicolumn{1}{|c}{0.47} & \multicolumn{1}{c}{0.14} & \multicolumn{1}{|c}{0.25} & \multicolumn{1}{c}{0.03} & \multicolumn{1}{|c}{0.46} & \multicolumn{1}{c}{0.13} & \multicolumn{1}{|c}{0.27} & \multicolumn{1}{c}{0.08} & \multicolumn{1}{|c}{0.49} & \multicolumn{1}{c}{0.38} \\
    
    ImplicitCell & \multicolumn{1}{|c}{\textbf{0.28}} & \multicolumn{1}{c}{\textbf{0.06}} & \multicolumn{1}{|c}{\textbf{0.41}} & \multicolumn{1}{c}{\textbf{0.10}} & \multicolumn{1}{|c}{\textbf{0.21}} & \multicolumn{1}{c}{\textbf{0.02}} & \multicolumn{1}{|c}{\textbf{0.51}} & \multicolumn{1}{c}{0.12} & \multicolumn{1}{|c}{\textbf{0.23}} & \multicolumn{1}{c}{0.08} & \multicolumn{1}{|c}{\textbf{0.36}} & \multicolumn{1}{c}{\textbf{0.27}} \\
    \bottomrule
    \multicolumn{8}{l}{*Note: The initial poses of ImplicitVol were from our tracking system.}
    
    \end{tabular}

\end{table*}

\label{sec::phantom_quant}

\section{Experimental Results}
\subsection{Phantom Dataset}

Fig. \ref{fig::qualitative_phantom} presents the qualitative reconstruction results of the phantom dataset, with 3D volumes rendered using 3D Slicer \cite{FEDOROV20121323}. For each result, we show both a 3D rendering and a longitudinal slice that intersects with the phantom's wire. This slice visualization enables intuitive evaluation of pose refinement quality, where a straighter reconstructed wire indicates more precise pose refinement. The LFE metric below each slice quantifies the wire's straightness, with lower values indicating better reconstruction quality.

Under light noise conditions, discrete methods (FDP and FDP-TVR) show visibly wavy and jagged wire reconstructions, and the studs of Lego bricks in red box are not well-defined. INR-based methods (INR and ImplicitVol) demonstrate improved wire straightness and more refined Lego surfaces, but still exhibit noticeable intermittent artifacts. ImplicitCell achieves the straightest wire and the clearest Lego brick surfaces, with significantly fewer artifacts. For heavy noise conditions, the wire and Lego bricks in FDP, FDP-TVR, and INR reconstructions are severely distorted, with visible stair-stepping artifacts. ImplicitVol shows improved wire straightness and reduced artifacts, but still exhibits some fragments in the 3D rendering. ImplicitCell produces the most robust reconstruction, with the straightest wire and more defined Lego brick surfaces, demonstrating its superior ability to preserve finer structure while achieving precise pose refinement. The LFE metric further confirms ImplicitCell's superior performance, with the lowest LFE values under both light and heavy noise conditions.

\begin{figure}[t]
    \centerline{\includegraphics[width=0.75\columnwidth]{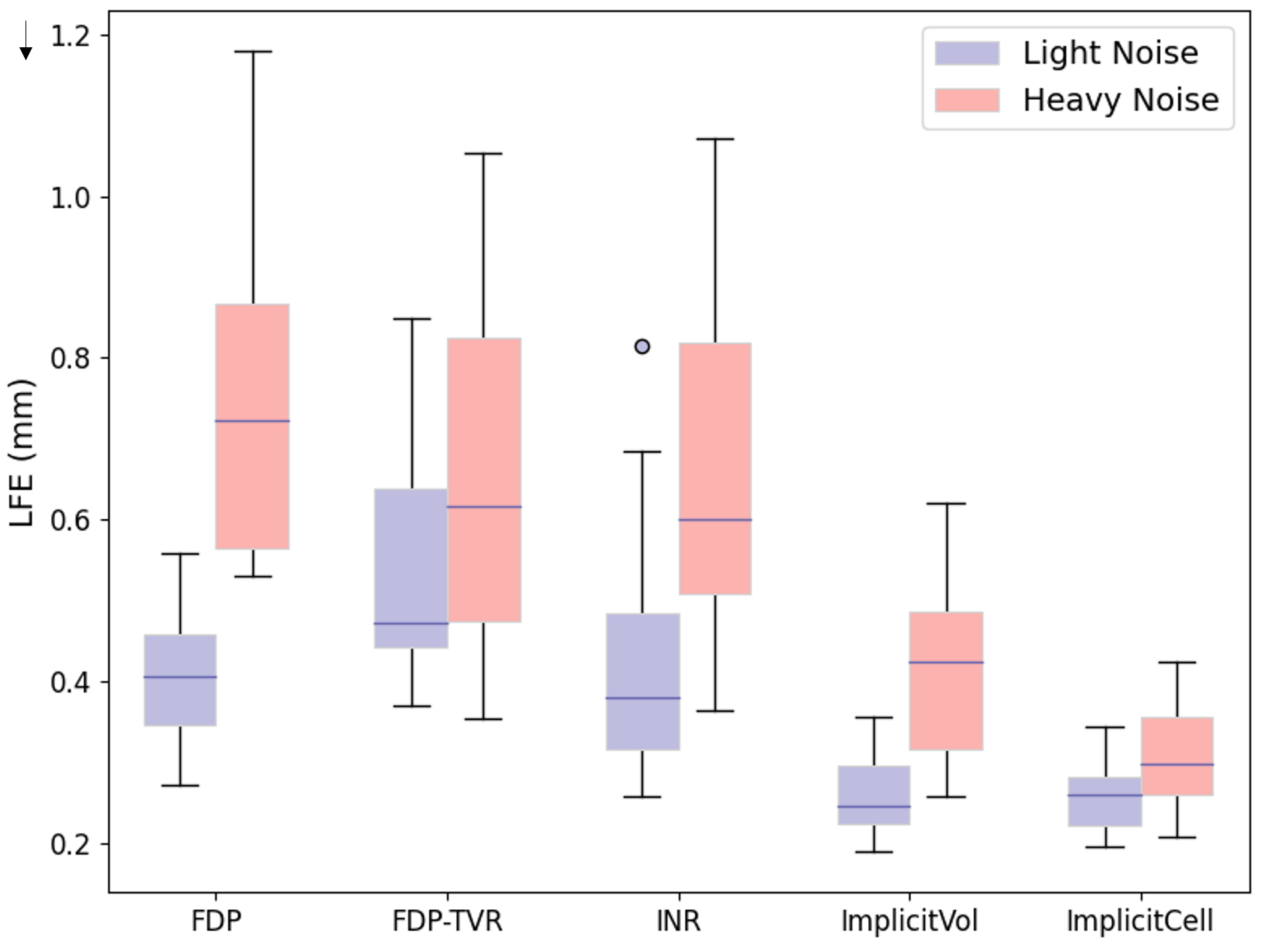}}
    \caption{Quantitative comparison of results on phantom datasets.} %
    \label{fig::quantitive_phantom}
\end{figure}

The quantitative results are shown in Fig. \ref{fig::quantitive_phantom}. Under both light and heavy noise conditions, ImplicitCell achieves the lowest median LFE with the smallest interquartile range (IQR), indicating its precision and robustness. 
The overall quantitative results across phantom dataset are demonstrated in Table \ref{tab:performance_results}. ImplicitCell demonstrates a substantial improvement compared to all baselines, and it outperforms traditional methods like FDP and FDP-TVR by roughly 50\%. Compared to the INR-based baselines, it outperforms INR by about 49\% and ImplicitVol by around 15\%.

\subsection{CCA Dataset}
\begin{figure*}[bp]
    \centerline{\includegraphics[width=1.99\columnwidth]{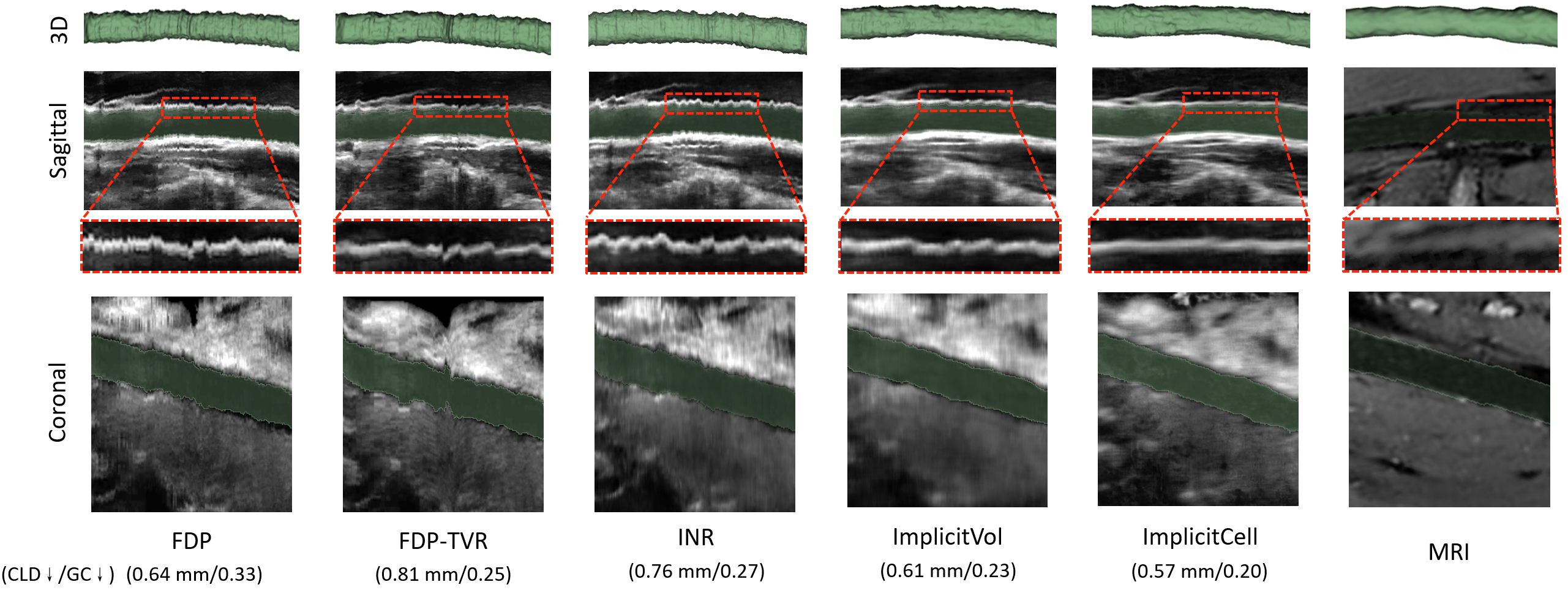}}
    \caption{Visual comparison of artery reconstruction quality across different methods and MRI reference on CCA dataset. For each method, the top row shows a 3D rendering of the artery segmentation. The middle and bottom rows display sagittal and coronal slices of the reconstructed volumes, respectively, with the artery segmentation overlaid in green. The red dashed boxes in the sagittal slices magnify the detail of artery wall, and the magnifying MRI image is enhanced for better visualization. Numerical values below each method represent ``(CLD/GC)" metrics. The lower value indicates the better result.}
    \label{fig::MRI}%
\end{figure*}

Fig. \ref{fig::MRI} visually compares the reconstruction quality of the CCA dataset across different methods, with the CCA segmentation highlighted in green. For each method, we present a 3D volume rendering of the CCA segmentation, along with sagittal and coronal slices of the reconstructed volumes, overlaid with the green CCA segmentation. The sagittal slices, magnified within red dashed boxes, are particularly useful for observing fine details of the vessel wall. The MRI image is included as the gold standard reference. The numerical values below each method represent the quantitative assessment of the reconstruction quality. Lower CLD indicates greater accuracy in aligning with the MRI reference, and lower GC suggests fewer artifacts and a smoother reconstructed surface. \newline
\begin{figure}[tp]
    \centering
    \centerline{\includegraphics[width=0.75\columnwidth]{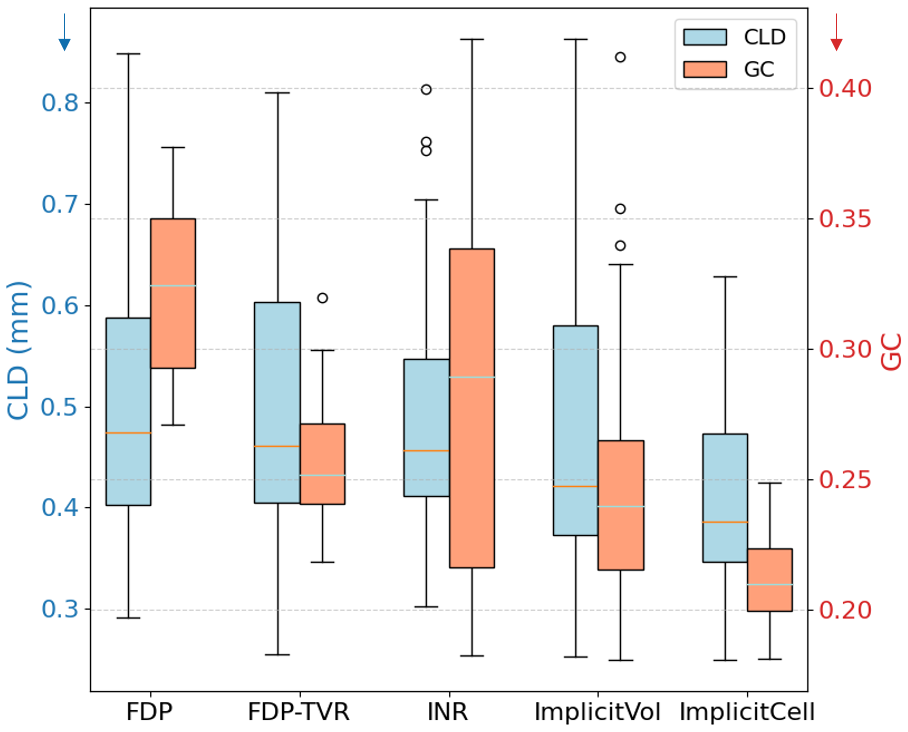}}
    \caption{Quantitative comparison of different methods on CCA datasets.}
    \label{fig::MRI_quant}
\end{figure}
\indent In 3D rendering, FDP, FPD-TVR and INR appear relatively rough and jagged surface, while ImplicitVol shows some improvement with less artifacts. ImplicitCell achieves the most natural appearance that closely resembles the MRI reference. In the sagittal slices, FDP and FDP-TVR exhibit stair-stepping artifacts along the vessel wall, and INR and ImplicitVol show modest improvements but still retain artifacts. ImplicitCell achieves superior results with clearer vessel structures and virtually eliminated artifacts, closely approximating the contour seen in the MRI reference. In the coronal slices, baseline methods show varying degrees of roughness and blurriness, while ImplicitCell produces crisp and well-defined vessel structures. The quantitative metrics below each method support the visual observations. ImplicitCell achieves the lowest CLD (0.57 mm) and GC (0.20), indicating the closest to MRI reference and highest surface smoothness, respectively. \newline
\indent The distributions of CLD and GC across different methods are presented as boxplots in Fig. \ref{fig::MRI_quant}.  For both CLD and GC, ImplicitCell achieves not only the lowest median values but also the narrowest IQRs compared to baselines, indicating more consistent reconstruction quality across different subjects. 
The quantitative results across all metrics demonstrate ImplicitCell's advantages over baseline methods, outperforming baselines by 10-20\% and 16-30\% respectively, as shown in Table \ref{tab:performance_results}.

\subsection{CA Dataset}

For visual evaluation of the CA dataset, we compared the aligned slices from reconstructed volumes against clinical longitudinal ultrasound reference scans (Fig. \ref{fig::clinical}). This enables direct assessment of plaque morphology, which is critical for accurate diagnosis and risk stratification. The quantitative metrics below each method provide numerical assessment of reconstruction quality through image similarity and plaque morphology measures. Lower LPIPS and higher SSIM indicate better image similarity, while lower $\Delta$SII suggests closer alignment with the reference on plaque morphology.
\setcounter{figure}{7}
\begin{figure*}[bp]
    \centerline{\includegraphics[width=1.7\columnwidth]{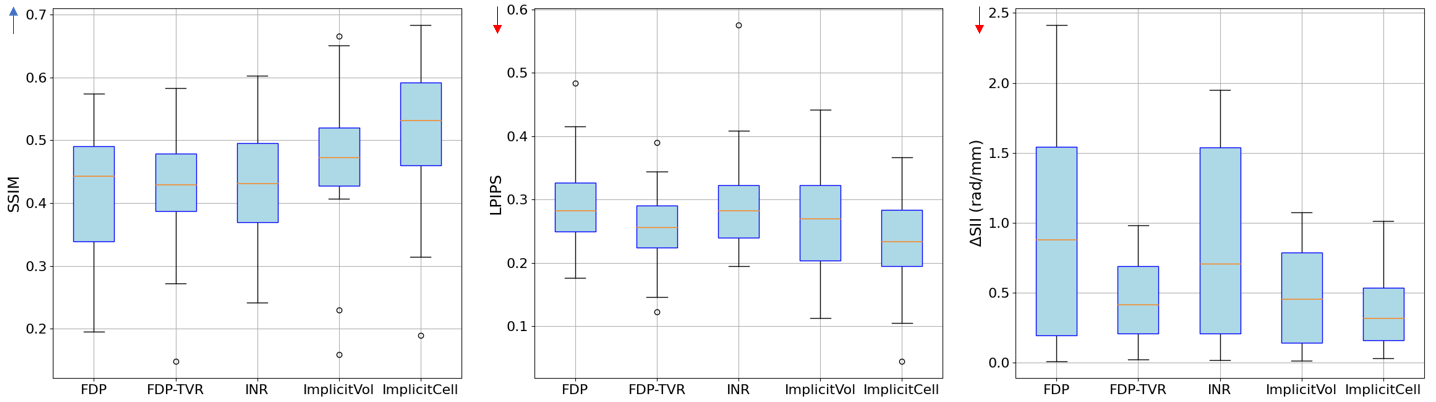}}
    \caption{Quantitative comparison of different methods on CA datasets.}
    \label{fig::clinical_quant}
\end{figure*}

As demonstrated in Fig. \ref{fig::clinical}, FDP, FDP-TVR and INR exhibit heavily degraded and poorly defined plaque structures, making confident assessment of plaque morphology challenging. ImplicitVol shows some improvement in delineating the plaque boundary and presenting better overall morphology, but still exhibits a degree of blurring. In contrast, ImplicitCell achieves the\hfill most\hfill clinically\hfill valuable\hfill plaque\hfill visualization\hfill among\hfill the

\setcounter{figure}{6}
\begin{figure}[H]
    \centerline{\includegraphics[width=0.85\columnwidth]{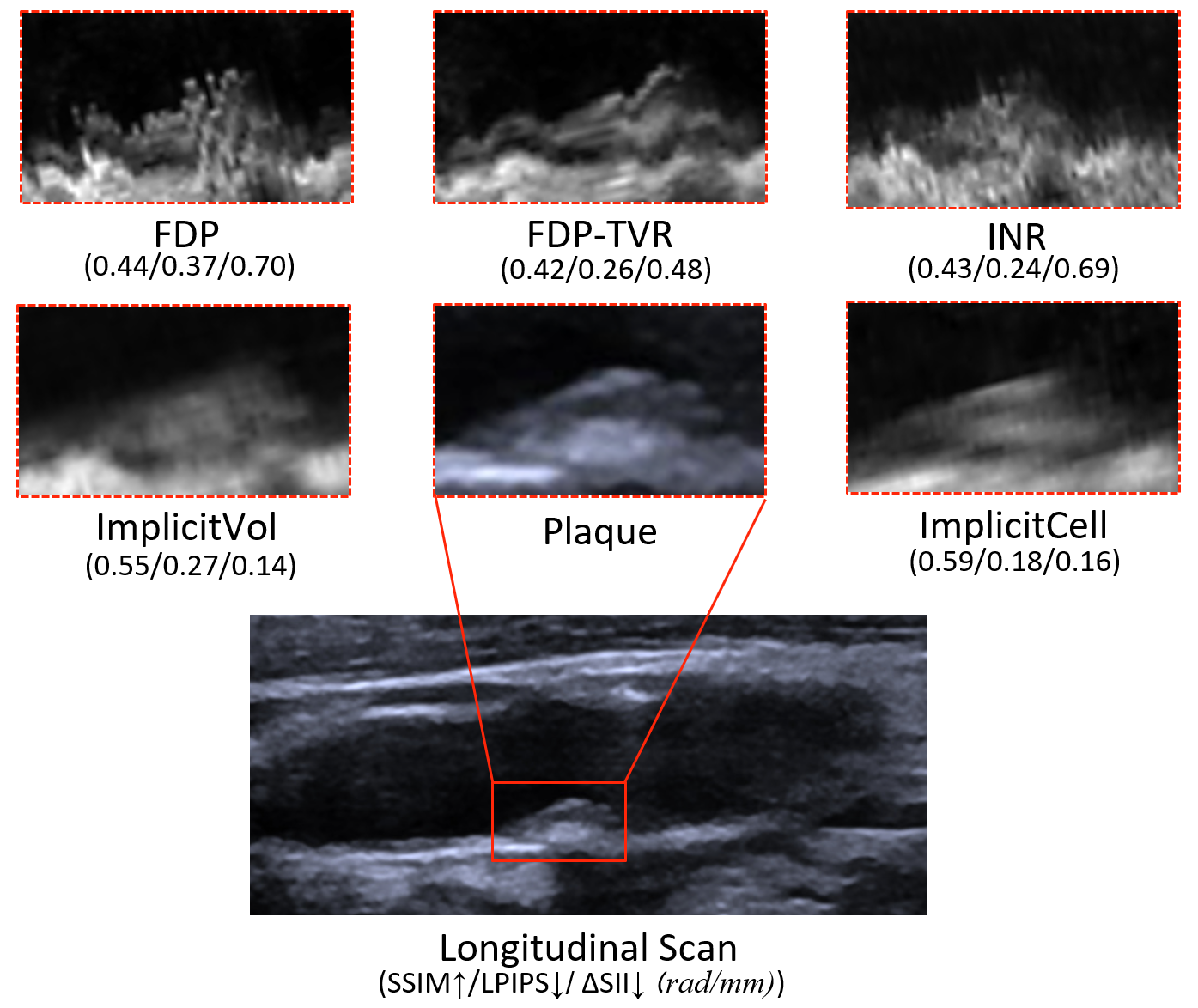}}
    \caption{The qualitative reconstruction results of the CA datasets. The slices of artery plaque in reconstructed volumes alongside longitudinal scans are dispalyed. The size of plaque is 8.7mm in length and 2.3mm in thickness. Numerical values below each method represent ``(SSIM/LPIPS/$\Delta$SII)" metrics.}
    \label{fig::clinical}
\end{figure}
\setcounter{figure}{8}
\noindent baseline methods, with well-defined boundaries and clear internal structure that closely match the morphology observed in the clinical longitudinal scan. Particularly, the internal plaque details (e.g., fibrous cap) in ImplicitCell reconstructions are the only ones that align well with the longitudinal US reference scan, demonstrating the clearest and most distinct visualization of the atherosclerotic plaque. The quantitative metrics below each method further confirm ImplicitCell's excellence, with the highest SSIM (0.41) and lowest LPIPS (0.26) indicating the finest reconstruction details, and the small $\Delta$SII (0.16 rad/mm) suggesting the resemblance with the reference in terms of plaque morphology.

The quantitative results in Fig. \ref{fig::clinical_quant} demonstrate ImplicitCell's leading performance across all metrics. For image similarity estimates, ImplicitCell achieves the highest SSIM and lowest LPIPS scores with smallest variance. ImplicitCell also shows better plaque morphology consistency through $\Delta$SII evaluation, demonstrating the narrowest IQR among all methods. As shown in Table \ref{tab:performance_results}, ImplicitCell outperforms baselines by around 10-25\% on SSIM and LPIPS, and 25-65\% on $\Delta$SII.

\subsection{Ablation Study} %

\begin{figure*}[t]
    \centerline{\includegraphics[width=1.9\columnwidth]{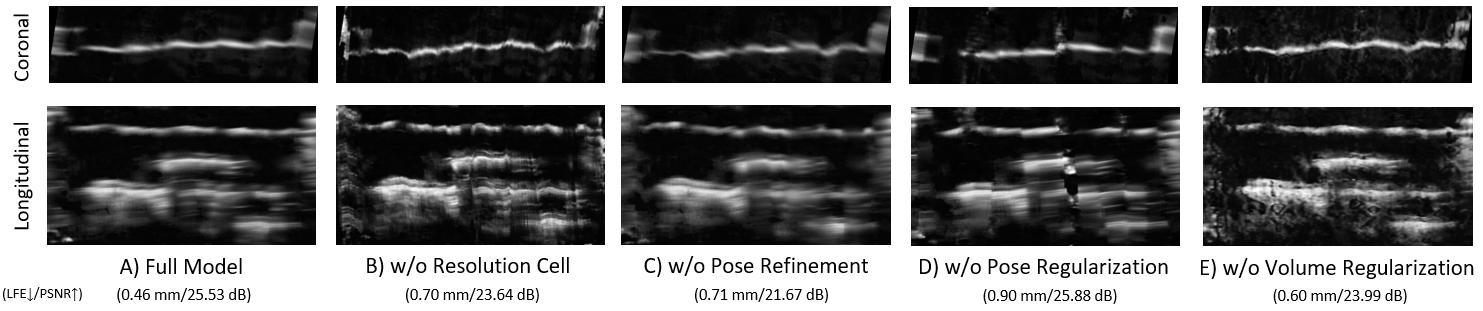}}
    \caption{Ablation results on phantom datasets with heavy noise.}
    \label{fig::ablation}
\end{figure*}

To assess the contribution of each component in ImplicitCell, we conducted ablation study on phantom dataset with heavy noise. We ablated the resolution cell model (Sec.~\ref{sec::cell_model}), pose refinement (Sec.~\ref{sec::pose_refinement}), pose regularization (Sec.~\ref{sec::pose_reg}) and volume regularization (Sec.~\ref{sec::volume_reg_}). The quantitative results are shown in Table~\ref{tab:ablation_results}, and the visual comparison is presented in Fig.~\ref{fig::ablation}. The full model demonstrates superior performance on both LFE and PSNR.

Ablating resolution cell model leads to substantial performance degradation, with LFE increasing by 68\% and PSNR dropping to 24.76±0.86 dB. As shown in Fig.~\ref{fig::ablation}B, the reconstructed volume becomes noisier with wavier wires, demonstrating that modeling overlapping resolution cells is crucial for accurate volume reconstruction and pose refinement.

Without pose refinement, reconstruction quality degrades significantly with PSNR dropping to 22.75±1.04 dB. Fig.~\ref{fig::ablation}C shows increased blurriness and reduced wire straightness, demonstrating that the performance of INR model is significantly affected by the quality of pose signals.

Ablating pose regularization results in the largest accuracy deterioration, with LFE increasing to 0.89±0.15 mm. As shown in Fig.~\ref{fig::ablation}D, despite maintaining comparable PSNR, the reconstructed volume exhibits significant artifacts and discontinuities, indicating that pose refinement optimization falls into local minima without regularization.

Without volume regularization, the reconstructed volume becomes noisier, with PSNR dropping to 24.30±1.34 dB. As shown in Fig.~\ref{fig::ablation}E, the volume exhibits increased noises and artifacts, highlighting the importance of volume regularization in maintaining reconstruction quality.
\begin{table}[tp]
    \centering
    \caption{Ablation Study Results}
    \label{tab:ablation_results}
    \begin{tabular}{l c c c c}
    \toprule
     & \multicolumn{2}{c}{$\downarrow$ LFE (\textit{mm})} & \multicolumn{2}{c}{$\uparrow$ PSNR (dB)} \\
    \cmidrule(r){2-3} \cmidrule(l){4-5}
    Method & \multicolumn{1}{c}{Mean} & \multicolumn{1}{c}{SD} & \multicolumn{1}{c}{Mean} & \multicolumn{1}{c}{SD} \\
    \midrule
    Full Model & \multicolumn{1}{|c}{0.38} & \multicolumn{1}{c}{0.06} & \multicolumn{1}{|c}{26.16} & \multicolumn{1}{c}{0.89} \\
    w/o Resolution Cell & \multicolumn{1}{|c}{0.64} & \multicolumn{1}{c}{0.19} & \multicolumn{1}{|c}{24.76} & \multicolumn{1}{c}{0.86} \\
    w/o Pose Refinement & \multicolumn{1}{|c}{0.51} & \multicolumn{1}{c}{0.19} & \multicolumn{1}{|c}{22.75} & \multicolumn{1}{c}{1.04} \\
    w/o Pose Regularization & \multicolumn{1}{|c}{0.89} & \multicolumn{1}{c}{0.15} & \multicolumn{1}{|c}{26.00} & \multicolumn{1}{c}{0.78} \\
    w/o Volume Regularization & \multicolumn{1}{|c}{0.38} & \multicolumn{1}{c}{0.11} & \multicolumn{1}{|c}{24.30} & \multicolumn{1}{c}{1.34} \\
    \bottomrule
    \end{tabular}
\end{table}
\section{Discussion and Conclusion} %
We propose ImplicitCell, a physics-aware framework for freehand 3DUS reconstruction that jointly optimizes volume reconstruction and pose refinement through INR and resolution cell modeling. Our comprehensive experiments on phantom and carotid datasets demonstrate significant improvements in reconstruction quality, particularly under challenging conditions with noisy EM tracking.

ImplicitCell's effectiveness stems from two key aspects: 1) the physics-based resolution cell modeling accurately captures spatial overlap between adjacent US pixels, providing strong geometric constraints for pose refinement, and 2) the joint optimization framework enables mutual enhancement between volume reconstruction and pose refinement through INR's continuous representation. This synergy allows ImplicitCell to achieve robust reconstruction even with noisy EM tracking data, as demonstrated in our comprehensive experiments on phantom and carotid datasets.

It is important to acknowledge two key limitations of ImplicitCell. Firstly, optimization robustness can be challenged when adjacent pose signal noise exceeds slice thickness along the longitudinal axis - in such cases combining ImplicitCell with methods like TVR may achieve more robust reconstruction. Secondly, global distortion handling is limited by local US image consistency, which may not fully resolve EM tracking distortions in metal-rich environments. %

In summary, ImplicitCell integrates an INR model with an ultrasound resolution cell model to enhance volume reconstruction and pose signal refinement for freehand 3DUS. Our method significantly improves volume accuracy and reduces artifacts, proving particularly effective in POCUS setups with noisy EM tracking systems. ImplicitCell enables reliable and high-quality 3DUS imaging on cost-effective systems, demonstrating broader potential for clinical applications.

\bibliographystyle{ieeetr}
\normalem
\bibliography{processed_ref}

\end{document}